\documentclass{article}
\usepackage{amsmath}
\usepackage{amsfonts}
\def\1ad{\mbox{\normalsize $^1$}}
\def\2ad{\mbox{\normalsize $^2$}}
\def\3ad{\mbox{\normalsize $^3$}}
\def\4ad{\mbox{\normalsize $^4$}}
\def\5ad{\mbox{\normalsize $^5$}}
\def\6ad{\mbox{\normalsize $^6$}}
\def\7ad{\mbox{\normalsize $^7$}}
\def\8ad{\mbox{\normalsize $^8$}}
\def\makefront{
\vspace*{1cm}\begin{center}
\def\sp{
\renewcommand{\thefootnote}{\fnsymbol{footnote}}
\footnote[4]{corresponding author : \email_speaker}
\renewcommand{\thefootnote}{\arabic{footnote}}
}
\def\newtitleline{\\ \vskip 5pt}
{\Large\bf\titleline}\\
\vskip 1truecm
{\large\bf\authors}\\
\vskip 5truemm
\addresses
\end{center}
\vskip 1truecm
{\bf Abstract:}
\abstracttext
\vskip 1truecm
}
\def\be{\begin{equation}}                      
\def\ee{\end{equation}}                        
\def\bea{\begin{eqnarray}}                     
\def\eea{\end{eqnarray}}
\def\nn{\nonumber}
\def\bm{\bibitem}
\def\ie{{\it i.e.\ }}

\def\im{{\rm i}}      

\def\oneone{\rlap 1\mkern4mu{\rm l}}
\def\demand{\buildrel!\over=}
\newcommand{\ft}[2]{{\textstyle{\frac{{\scriptstyle #1}}{\scriptstyle
#2}}}}

\newcommand{\ZZ}{\mathbb{Z}}

\newcommand{\RR}{\mathbb{R}}

\begin {document}                 
\rightline{Imperial/TP/02-03/13}
\rightline{hep-th/0302164}
\def\titleline{Form-field Gauge Symmetry in M--theory}
\def\email_speaker{
{\tt 
k.stelle@imperial.ac.uk}}
\def\authors{J.~Kalkkinen\1ad and K.S.~Stelle\2ad\sp}
\def\addresses{\1ad Institut des Hautes \'Etudes Scientifiques, Le Bois-Marie\\
35, Route de Chartres, Bures-sur-Yvette F-91440, France\\
\2ad The Blackett Laboratory, Imperial College, London SW7 2AZ, UK}
\def\abstracttext{We show how to cast an interacting system of M--branes into manifestly
gauge-invariant form using an arrangement of higher-dimensional Dirac surfaces.
Classical M--theory has a cohomologically nontrivial and noncommutative set of gauge
symmetries when written using a ``doubled'' formalism containing 3-form and 6-form gauge
fields. We show how the arrangement of Dirac surfaces allows an integral subgroup of these
symmetries to be preserved at the quantum level. The proper
context for discussing these large gauge transformations is
relative cohomology, in which the 3-form transformation parameters
become exact when restricted to the five-brane worldvolume. This structure yields the
correct lattice of M-theory brane charges. 
}
\large
\makefront

\section{Introduction}

The bosonic sector of $D=11$ supergravity is derived from the action
\be
I^{\rm bulk}_{11} = \int_X(R\ast\oneone -\ft12G_4\wedge \ast G_4-\ft16C_3\wedge G_4\wedge
G_4)\ ,
\ee
where $G_4=dC_3$ and the relative coefficients are fixed by the requirement of $D=11$
local supersymmetry once the fermions have been included. The $D=11$ spacetime $X$ is
taken here to be without boundary.

The form-field $C_3$ has the field equation
\be
d\ast G_4+\ft12 G_4\wedge G_4 = 0\ ;
\ee
this is manifestly invariant under the gauge transformation $\delta C_3 = \Lambda_3$,
where $d\Lambda_3=0$. This allows ``large'' gauge transformation if $\Lambda_3$ is
taken to be closed but not exact; $\delta C_3$ is ``small'' if
$\Lambda_3=d\lambda_2$, for $\lambda_2$ globally defined.

Rewriting the $C_3$ field equation as $d(\ast G_4+\ft12 C_3\wedge G_4)=0$, notice that
one can introduce a dual field strength
\be 
\tilde G_7 = d\tilde C_6-\ft12 C_3\wedge G_4
\ee
and impose the duality condition 
\be
\tilde G_7=\ast G_4\ ;
\ee
then the $C_3$ field equation becomes a Bianchi identity:
\be
d\tilde G_7 = -\ft12 G_4\wedge G_4 .
\ee
In this way, the $C_3$ field equation may be replaced by a duality condition for the
``doubled'' $(C_3,\tilde C_6)$ system.

This doubled system has a noncommutative ring of large gauge transformations:
\bea
\delta C_3 =\Lambda_3\ ,\quad&&\quad\delta\tilde C_6 = \Lambda_6-\ft12\Lambda_3\wedge
C_3\cr
[\delta_{\Lambda_3},\delta_{\Lambda'_3}] = \delta_{\Lambda_6}\quad
&\makebox[0pt]{with}&\
\quad
\Lambda_6=\Lambda_3\wedge\Lambda'_3\cr
[\delta_{\Lambda_3},\delta_{\Lambda_6}] &=& 
[\delta_{\Lambda_6},\delta_{\Lambda'_6}] =0\ .\label{doubledalg}
\eea
This is the cohomology ring for 3-forms and 6-forms on the underlying spacetime.
These cohomologies are taken for the time being to be defined over the real numbers, but
they will soon be restricted to integral cohomologies when we consider the corresponding
Dirac quantization conditions.

In the rest of this article, which is based on \cite{mbranealg}, we will investigate the
way in which the algebra (\ref{doubledalg}) is preserved {\it a)} in the presence of
2--branes and 5--branes, and {\it b)} at the quantum level.

\section{Current couplings to 2--branes and 5--branes}

For an ${\rm M}_2$--brane worldvolume $W_3$ ending on an ${\rm M}_5$--brane worldvolume
$W_6$, one has $\partial W_3\ne0$, so the basic ${\rm M}_2$ coupling $\int_{W_3}C_3$
fails to be gauge invariant even for small gauge transformations $\Lambda_3=d\lambda_2$.
The cure for this problem is provided by a form-field that exists on the $W_6$
worldvolume: the self-dual 3--form $h_3$, which has a potential $b_2$. Using the latter,
one can take the combination $\int_{W_3}C_3 - \int_{\partial W_3}b_2$, which is invariant
under small gauge transformations when taken together with a compensating Green-Schwarz
mechanism, $\delta C_3=d\lambda_2$, $\delta B_2=\lambda_2$.

Note that a self-dual 3--form field strength is precisely what is needed in order to
complete the bosonic part of the $(2,0)$ worldvolume fluctuation-field supermultiplet.
The transverse oscillations of the 5--brane provide 5 worldvolume scalar bosonic degrees
of freedom, while the 16 broken supersymmetries contribute 8 worldvolume fermionic
degrees of freedom (taking into account that the fermionic equations of motion are of
first-order). Thus, in order to have a bose--fermi balance on the $W_6$ worldvolume, one
needs to have an additional 3 bosonic degrees of freedom. This is what is contributed by
the self-dual 3-form, which contributes precisely $\ft12(4\cdot3/2)=3$ degrees of
freedom.

The gauge-invariant field strength for the $C_3$ gauge field is accordingly the bulk
$\oplus$ worldvolume combination
\be
h_3 = i^\ast C-db_2
\ee
where $i^\ast$ is the pullback to the $W_6$ worldvolume effected by the $i:
W_6\hookrightarrow X$ embedding map.

The action for the ${\rm M}_2$, ${\rm M}_5$ system \cite{m2m5refs} can be written
\be
I_{\rm branes}=I_{\rm kinetic}+I^{\rm brane}_{\rm forms}+I_{\rm WZ}+I_{\rm counterterms}
\ee
in which the various terms are
\bea
I_{\rm kinetic} &=&
T_3\int_{W_3}d^3\xi\sqrt{-\det(\gamma_{\mu\nu})} +
T_6\int_{W_6}\sqrt{-\det(G_{ij})}\label{genaction1}\\
I^{\rm brane}_{\rm forms} &=& \ft12\int_{W_6}h\wedge\ast h + b\int_{W_6}\tilde C +
e\int_{W_6}h\wedge C + a\int_{W_3}C +k\int_{W_2}b\label{genaction2}\\
I_{\rm WZ} &=& f \int_{W_7}C\wedge G + \frac{q}2\int_{W_8}G\wedge G\ ;\label{genaction3}
\eea
the values of the coefficients $a,b,e,k,f,q$ shall be determined by requiring gauge
invariance. The selection of possible terms in (\ref{genaction1}--\ref{genaction3}) is
made by taking all the relevant products of operators, integrated over spaces of
appropriate dimensionality. Clearly, the $I_{\rm WZ}$ terms are unusual, and their
inclusion will need to be explained. The final term, $I_{\rm counterterms}$, is required
for anomaly cancellation and will involve both worldvolume and bulk contributions.

In order to determine the values of the coefficients in
(\ref{genaction1}--\ref{genaction3}), one has to impose the requirements of gauge
invariance but also take into account the fact that the ``magnetically'' charged 5--brane
and the string boundary of the 2--brane give rise to violations of the normal Bianchi
identities for the corresponding bulk and worldvolume form-fields.

\section{Relative homology and cohomology, Bianchi identities and gauge invariance}

In order to express the violated Bianchi identities compactly, it is convenient to
use the language of relative cohomology.\footnote{For some original
applications of relative cohomology to branes, see
\cite{Stanciu:2000fz,Figueroa-O'Farrill:2001kz}.} Consider a pair of form-fields of
adjacent rank, $(C_k,C_{k-1})$. The first element in a pair is taken to be valued in the
bulk spacetime, but the second element is taken to be valued in the subspace with respect
to which the relative cohomology is being defined, in this case the 5--brane
worldvolume $W_6$. Using the pull-back $i^\ast$ from the bulk spacetime to
the worldvolume $W_6$ as above, one can define the relative exterior derivative (or
coboundary operator) as
\be
d(C_k,C_{k-1}):=(dC_k,i^\ast C_k-dC_{k-1})\ .\label{relcohomology}
\ee
The group of forms closed in this sense, taken for now over the real numbers,
is denoted $H^k(X,W_6;\RR)$.

The pair of field strengths $(G_4,h_3)$, valued respectively in the bulk and in the
worldvolume $W_6$, is thus given locally in terms of the exterior derivative
(\ref{relcohomology}): $(G_4,h_3)=d(C_3,b_2)$, so the na\"{\i}ve Bianchi identity is
$d(G_4,h_3)=0$. In the presence of magnetically charged sources, however, this must be
violated on the corresponding source loci:
\be
d(G_4,h_3) = (\kappa T_6\delta(W_6),T_{2\hookrightarrow 6}\delta(W_2))\
,\label{violatedBianchi}
\ee
where $T_6$ and $T_{2\hookrightarrow 6}$ are real coefficients; $\kappa$ is the
gravitational coupling constant, needed on dimensional grounds. $T_6$ has
been chosen to equal the 5--brane tension, as is necessary in order for the 5--brane to be
a $\ft12$ supersymmetric BPS soliton.

After some analysis \cite{mbranealg}, the various conditions for small gauge invariance,
taken together with the form (\ref{violatedBianchi}) of the violated Bianchi identities
plus use of the field equations yield the following relationships between the
coefficients in (\ref{genaction1}--\ref{genaction3}):
\bea
T_3 &=& -a = 2k\label{coefrels1}\\
T_6 &=& -2e = 6f\label{coefrels2}\\
T_{2\hookrightarrow 6} &=& -\frac{k}e = \frac{T_3}{T_6}\\
b &=& 0\ .\label{coefrels3}
\eea

These relations are fully consistent with the ``brane surgery'' relations
\cite{townsend,papadopoulos} expected on the basis of charge conservation for a $(q-1)$
brane intersecting a $(p-1)$ brane over a $(k-1)$ brane,
\be
T_{k\hookrightarrow q}T_q = T_{k\hookrightarrow p}T_p\ ,\label{branesurgery}
\ee
where $T_{k\hookrightarrow q}$ and $T_{k\hookrightarrow p}$ represent the tension of the
$(k-1)$ brane as seen within the $(q-1)$ brane or as seen within the $(p-1)$
brane. The relations (\ref{coefrels1}--\ref{coefrels3}) fit this rule for
$T_{2\hookrightarrow 6}=T_3/T_6$ if one takes $T_{2\hookrightarrow 3}=1$.

In addition to the coefficient relations (\ref{coefrels1}--\ref{coefrels3}), one obtains
also the following homology relations from the gauge invariance requirements:
\be
W_2 = \partial W_3\ ,\qquad W_6 = \partial W_7\ ;
\ee
the first of these is of course expected since the string is always located at the
boundary of the 2--brane on the 5--brane; the second establishes $W_7$ as a ``Dirac
surface'' for the 5--brane worldsheet $W_6$.

Homology relations are of course dual to cohomology relations, and so one has an
appropriate boundary relation that is dual to the relative exterior
derivative/coboundary operator defined in (\ref{relcohomology}). For cycles
$(W_k,W_{k-1})$ in
$(X,W_6)$ one has the relative boundary relation
\be
\partial(W_k,W_{k-1});=(W_{k-1}-\partial W_k,\partial W_{k-1})\ .\label{relhomology}
\ee
Thus in relative homology, the statement that a pair has no boundary means
\be
\partial(W_k,W_{k-1})=0 \Rightarrow W_{k-1}=\partial W_k\subset W_6\ ,\quad 
\partial W_{k-1}=0\ .\label{noboundary}
\ee
The group of homology chains in this sense, again taken for the time being over the reals,
is denoted $H_k(X,W_6;\RR)$. 

Pairs of cycles and pairs of forms can be integrated as follows:
\be
\int_{(W_k,W_{k-1})}(C_k,C_{k-1}) := \int_{W_k}C_k - \int_{W_{k-1}}C_{k-1}\
.\label{pairing}
\ee
The duality of relative homology and cohomology is then expressed {\it via} Stokes'
Theorem:
\be
\int_{(D_{k+1},D_k)}d(C_k,C_{k-1}) = - \int_{\partial(D_{k+1},D_k)}(C_k,C_{k-1})\
.\label{stokes}
\ee
One can also give the following meaning to integrals over linear combinations of spaces:
\be
\int_{\alpha W+\beta U}C := \alpha\int_W C + \beta\int_U C\ ,\alpha,\beta\in\RR\
.\label{surfacesum}
\ee

Relative cohomology language can now be used to describe the sense in which the gauge
transformations (\ref{doubledalg}) remain ``large'' in the presence of 2--branes and
5--branes. The transformation parameters $(\Lambda_3,\lambda_2)$ for the form-fields
$(C_3,b_2)$ are taken to be elements of $H^3(X,W_6;\RR)$. Thus, $\Lambda_3$ may remain
cohomologically nontrivial on $X-W_6$, but it must reduce to an exact form $d\lambda_2$
when restricted to the 5--brane worldvolume $W_6$.

\section{Dirac-Schwinger-Zwanziger quantization relations}

The above real relative homology and cohomology groups become restricted to integral
subgroups when quantum effects are taken into account. The basic requirement is that
adiabatic deformations of a dual pair of electric and magnetic solitons through a closed
deformation path should not produce any change in the quantum generating functional
path-integral. Thus, variations of the action $I$ by amounts $2\pi k$, $k\in \ZZ$ are
allowed since the integrand $\exp(\im I)$ becomes multiplied in that case just by
$\exp(2\pi\im k)=1$. A Wu-Yang style argument \cite{mbranealg} then shows that if an
${\rm M}_2$ brane worldvolume $W_3$ is deformed through a closed path $\Sigma_4$ around an
${\rm M}_5$ brane worldvolume $W_6$, one must for quantum consistency have at most a
phase change
\be
T_3\int_{W_3^{\rm final}}C - T_3\int_{W_3^{\rm initial}}C = T_3\int_{\Sigma_4}G \demand
2\pi\ell\ ,\quad\ell\in\ZZ\ ,
\ee
\ie the class 
\be
{T_3\over2\pi}\left[G_4\right]\label{g4integral}
\ee 
must be integral when integrated over closed
manifolds $\Sigma_4$ that do not intersect the 5--brane worldvolume $W_6$ itself.

Taking then $\Sigma_4=\partial D_5$ and using the violated Bianchi identity $dG=\kappa
T_6\delta(W_6)$ yields the ${\rm M}_2$-${\rm M}_5$ brane Dirac quantization rule for the
cell units of the brane charge/tension lattice:
\be
\kappa T_3T_6 \demand 2\pi\ .\label{diracrel}
\ee

In addition, one has a quantization relation for the self-dual (\ie dyonic) string on the
5--brane worldvolume $W_6$. Firstly, note that the dimension here is in the
sequence $d=4n+2$, $n\in \ZZ$, for which the Dirac-Schwinger-Zwanziger quantization
condition for dyons with (electric,magnetic) charges $(e_i,g_i)$ is {\em symmetric}
\cite{bremer}:
\be
e_1g_2 + e_2g_1 = 2\pi \ell\ ,\quad\ell\in\ZZ\ .\label{dszrel}
\ee
Thus, for dyons with $e=g$, one has the charge relation $e_1e_2=\pi \ell$,
$\ell\in\ZZ$. This might appear to give a unit cell that is out by a factor of $\ft12$
with respect to the unit expected, but one needs to recall that the coefficient $k$ of
$\int_{W_2}b$ in (\ref{genaction2}) also contained a factor of $\ft12$ in the
coefficient relations (\ref{coefrels1}). Taking this factor into account and performing
an adiabatic closed deformation similar to those above, one obtains
\be
T_3\int_{(W_3,W_2)^{\rm final}}(C_3,b_2) - T_3\int_{(W_3,W_2)^{\rm initial}}(C_3,b_2) =
T_3\int_{(\Sigma_4,\Sigma_3)}(G_4,h_3) \demand 2\pi\ell\ ,\ell\in\ZZ\ .
\ee
Then taking $(\Sigma_4,\Sigma_3)=\partial(D_5,D_4)$ and using the violated Bianchi
identities (\ref{violatedBianchi}), one finds out that of the two terms, only the $D_4$
integral contributes because the $D_5$ integral of $\delta(W_6)$ vanishes since
$\hbox{dim}(D_5\cap W_6^\perp)=1$ only. Thus one obtains
$T_3\int_{(\Sigma_4,\Sigma_3)}(G_4,h_3) = T_3\int_{D_4}T_{2\hookrightarrow 6}\delta(W_6)$
and accordingly there is a second quantization rule for the charge/tension lattice-cell
units:
\be
T_3T_{2\hookrightarrow 6} \demand 2\pi\ .\label{stringdiracrel}
\ee
Combining this with the relation (\ref{coefrels3}) for $T_{2\hookrightarrow 6}$, one
obtains the quadratic cell unit rule
\be
(T_3)^2 \demand 2\pi T_6\ ,\label{quadraticrule}
\ee
The quantization rules (\ref{diracrel}) and (\ref{quadraticrule}) yield the correct
charge lattice for M--theory solitons
\cite{Schwarz:1996jq,deAlwis:1996ez,deAlwis:1997gq,Lavrinenko:1999xi,Julia:2000af}.

\section{$D=12$ Formulation and Dirac Surfaces}

The bulk plus brane-source action discussed so far for M--theory needs to be
completed by gravitational counterterms in order to cancel diffeomorphism anomalies on
the 5--brane worldvolume $W_6$ that arise from loops of worldvolume chiral fermion modes
\cite{Freed:1998tg}. In order to write these, it is convenient to introduce a $D=12$
spacetime $Y$ such that
\be
X=\partial Y\ .\label{elevenboundary}
\ee
Using this, the Chern-Simons term in the bulk action can be conveniently written
$-\frac1{6\kappa}\int_Y G_4\wedge G_4\wedge G_4$. 

The main advantage of the $D=12$
formulation, however, is in the way it gives to express the Dirac surfaces needed to
maintain manifest gauge invariance. For this purpose, we need to introduce two surfaces
bounded by
$W_6$:
$V_7$, which extends into $Y$ in such a way that 
\be
i^\ast\delta_Y(V_7) = \delta_X(W_6)\ ,
\ee
and another surface $W_7\subset X$; for both one has the boundary relation
\be
\partial V_7=\partial W_7 =W_6\ .\label{w6boundaryrels}
\ee
Since $V_7$ and $W_7$ share a boundary, one may flip the orientation of $V_7$ and glue it
onto $W_7$ in order to make a closed surface $W_7\cup(-V_7)=W_7-V_7$\,:
\be
\partial(W_7-V_7)=W_6-W_6=0\ .
\ee
Hence, one can find a ball $V_8\subset Y$ such that
\be
\partial V_8 = W_7-V_7\ .
\ee

A similar construction can be made on the 5--brane worldvolume for surfaces bounded by
$W_2$. Letting $i: W_6\hookrightarrow W_7$, introduce $U_3\subset W_7$ with $\partial
U_3 = W_2$ such that 
\be
i^\ast\delta_{W_7}(U_3) = \delta_{W_6}(W_2)\ .
\ee
Since also one has $\partial W_3=W_2$, one can flip the orientation of $W_3$ and glue it
onto $U_3$ along $W_2$ to produce a closed surface $U_3\cup(-W_3)=U_3-W_3$ which can
similarly be taken to be the boundary of a ball $U_4$\,:
\be
\partial U_4 = U_3-W_3\ .
\ee

Taken all together, one has the following relative homology relations for the Dirac
surfaces:
\bea
\partial(V_8,W_7) &=& (V_7,W_6)\cr
\partial(U_4,U_3) &=& (W_3,W_2)\ .\label{diracboundaryrels}
\eea

At this point, we can also state the gauge invariance requirement for the integration
domain in the remaining term $\frac{q}2\int_{W_8}G_4\wedge G_4$ in the action
(\ref{genaction3}), with a coefficient to be understood in the sense of Eq.\
(\ref{surfacesum}):
\be
qW_8=-\frac{T_6}3V_8\ .\label{domainrel}
\ee

Taken all together, the gauge-field part of the action is then
\bea
 I_{\rm gauge} &=& 
       \int_X \frac{1}{2\kappa}~ G_4 \wedge * G_4 
     + {T_3} G_4 \wedge \Omega_7(R) 
     - \int_Y \frac{1}{6\kappa} G_4 \wedge G_4 \wedge G_4 
       \nn\\
& &  - \ft12 T_6 \int_{(V_8,W_7)} 
       \Big(  G_4 \wedge G_4 - 2\kappa T_3 ~\Omega_8, 
               ~ h_3 \wedge i^\ast G_4 \Big)    
       \label{final} \\
& &  + \frac{T_6}{4}  \int_{W_6} h_3\wedge \ast h_3
     + {T_3}  \int_{U_4} G_4 
     - \frac{T_3}{2} \int_{U_3} h_3 \ ,\nn
\eea
where the integration domains satisfy the
relations (\ref{elevenboundary},\ref{diracboundaryrels}). The terms involving $\Omega_8$
and $\Omega_7$ are parts of the counterterm structure needed to cancel the diffeomorphism
anomalies on the $W_6$ worldvolume. $\Omega_8=d\Omega_7$ generates the anomaly compensator
${\cal A}_6$ by transgression: 
\be
\delta_{\rm diff}\Omega_7 = d{\cal A}_6\ .
\ee
${\cal A}_6$ then cancels a part of the anomaly from loops of chiral fermions and of the
self-dual $h_3$ field in the 5--brane's $W_6$ 
$(2,0)$ supersymmetric worldvolume theory.

\section{Dependence on Dirac surfaces}

The presence of terms like $\int_{V_8}G_4\wedge G_4$ in the action (\ref{final}) may give
rise to concern whether the action as presented describes correctly the 2--brane/5--brane
system, or whether extra degrees of freedom have sneaked in {\it via} the dynamics of
surfaces like $V_8$.

The answer is ``no.'' One has just the required degrees of freedom and nothing more. This
is demonstrated by showing that variations of the Dirac surfaces in the action
(\ref{final}) produce effects that vanish modulo
$2\pi$ as a result of appropriate integrality and relative homology conditions.

Independence from variations of $Y$ and $V_8$ under shifts by closed surfaces boils down
to requiring that the classes
\be
{1\over2\pi\kappa}\left[{1\over
3!}G_4^3\right]\qquad\hbox{and}\qquad{T_6\over2\pi}\left[{1\over2!}G_4^2\right]
\ee
be integral. From the flux integrality condition (\ref{g4integral}) that 
\be
\left[{T_3\over2\pi}G_4\right] \in H^4(Y;\ZZ)\ ,\label{g4integralgroup}
\ee 
one sees that the needed integrality
conditions follow from the charge-lattice unit conditions
\be
{(2\pi)^2\over\kappa} = (T_3)^3 \qquad\hbox{and}\qquad 2\pi T_6 = (T_3)^2\ ,
\ee
which follow from (\ref{diracrel},\ref{quadraticrule}).

Independence from variations of $W_7$, $V_7$, $U_4$ and $U_3$ is more complicated because
some of these surfaces are subsurfaces of others that can be varied, and so are carried
along. Thus, varying $Y$ and $W_7$ by closed surfaces $\partial Z$ and $\partial D_8$, one
induces variations
\bea
Y'=Y+\partial Z \quad&\Rightarrow&\quad V_7'=V_7+\partial Z_8\cr
W_7'=W_7+\partial D_8 \quad&\Rightarrow&\quad U_3'=U_3+\partial D_4\ .
\eea
At the same time, one should also consider the variations
\bea
V_8' &=& V_8 + T_8\cr
U_4' &=& U_4 + T_4\ .
\eea

The shift in the action (\ref{final}) under these Dirac surface variations is
\cite{mbranealg}
\be
{T_6\over2}\int_{D_8-T_8-Z_8}G_4\wedge G_4 + 2\pi\int_{T_8}\Omega_8 + T_3\int_{T_4-D_4}G_4
+ {T_3\over2}\int_{(D_4,\partial D_4)}(G_4,h_3)\ .
\ee
Using again the flux integrality condition (\ref{g4integralgroup}), one finds that this
shift is in $2\pi\ZZ$ provided one has the boundary conditions for the integration domains
\be
\partial(-D_8+T_8+Z_8)=\partial T_8=\partial(T_4-D_4)=0\ ,
\ee
which follow from the Dirac surface relative homology relations (\ref{diracboundaryrels}).
In addition, one needs to require that 
\be
\left[\Omega_8\right]\in H^8(Y;\ZZ)\ .
\ee
This condition is known to be required also by membrane tadpole cancellation requirements
\cite{Sethi:1996es,Witten:1997md}.

\section{Lattice of Large Gauge Transformations}

Finally, we return to the large gauge transformations. Since the various Dirac and DSZ
quantization conditions restrict the M--theory charges to lie on the charge lattice
determined by (\ref{diracrel},\ref{stringdiracrel},\ref{quadraticrule}), the large gauge
transformations are also restricted. From the flux integrality condition
(\ref{g4integralgroup}), it follows that gauge transformations relating the gauge fields
on different hemispheres must also lie on an integral lattice,
\be
\left[{T_3\over2\pi}\Lambda_3\right]\in H^3(Y;\ZZ)\ .
\ee
Similarly, the flux integrality condition
\be
\left[{T_3\over2\pi}(G_4,h_3)\right]\in H^4(Y,W_6;\ZZ)
\ee
requires the gauge transformation integrality condition
\be
\left[{T_3\over2\pi}(\Lambda_3,\lambda_2)\right]\in H^3(Y,W_6;\ZZ)
\ee
and for the 6-form transformations one likewise finds the requirement
\be
\left[{T_6\over2\pi}\Lambda_6\right]\in H^6(Y;\ZZ)\ .
\ee
\section*{Remembrance} 
The work of Ref.\ \cite{mbranealg} on which this article is based was significantly aided
by early penetrating discussions with Sonia Stanciu, who is sadly no longer with us. Her
gentleness and her keen intelligence will be much missed.

\end{document}